\documentstyle[twoside,epsfig]{article}

\catcode`\@=11
\long\def\@makefntext#1{
\protect\noindent \hbox to 3.2pt {\hskip-.9pt  
$^{{\eightrm\@thefnmark}}$\hfil}#1\hfill}		

\def\@makefnmark{\hbox to 0pt{$^{\@thefnmark}$\hss}}	
	
\def\ps@myheadings{\let\@mkboth\@gobbletwo
\def\@oddhead{\hbox{}
\rightmark\hfil\eightrm\thepage}   
\def\@oddfoot{}\def\@evenhead{\eightrm\thepage\hfil
\leftmark\hbox{}}\def\@evenfoot{}
\def\sectionmark##1{}\def\subsectionmark##1{}}

\oddsidemargin=\evensidemargin
\newcounter{sectionc}\newcounter{subsectionc}\newcounter{subsubsectionc}
\renewcommand{\section}[1] {\vspace{12pt}\addtocounter{sectionc}{1} 
\setcounter{subsectionc}{0}\setcounter{subsubsectionc}{0}\noindent 
	{\tenbf\thesectionc. #1}\par\vspace{5pt}}
\renewcommand{\subsection}[1] {\vspace{12pt}\addtocounter{subsectionc}{1} 
	\setcounter{subsubsectionc}{0}\noindent 
	{\bf\thesectionc.\thesubsectionc. {\kern1pt \bfit #1}}\par\vspace{5pt}}
\renewcommand{\subsubsection}[1] {\vspace{12pt}\addtocounter{subsubsectionc}{1}
	\noindent{\tenrm\thesectionc.\thesubsectionc.\thesubsubsectionc.
	{\kern1pt \tenit #1}}\par\vspace{5pt}}
\newcommand{\nonumsection}[1] {\vspace{12pt}\noindent{\tenbf #1}
	\par\vspace{5pt}}
\newcounter{appendixc}
\newcounter{subappendixc}[appendixc]
\newcounter{subsubappendixc}[subappendixc]
\renewcommand{\thesubappendixc}{\Alph{appendixc}.\arabic{subappendixc}}
\renewcommand{\thesubsubappendixc}
	{\Alph{appendixc}.\arabic{subappendixc}.\arabic{subsubappendixc}}
\renewcommand{\appendix}[1] {\vspace{12pt}
        \refstepcounter{appendixc}
        \setcounter{figure}{0}
        \setcounter{table}{0}
        \setcounter{lemma}{0}
        \setcounter{theorem}{0}
        \setcounter{corollary}{0}
        \setcounter{definition}{0}
        \setcounter{equation}{0}
        \renewcommand{\thefigure}{\Alph{appendixc}.\arabic{figure}}
        \renewcommand{\thetable}{\Alph{appendixc}.\arabic{table}}
        \renewcommand{\theappendixc}{\Alph{appendixc}}
        \renewcommand{\thelemma}{\Alph{appendixc}.\arabic{lemma}}
        \renewcommand{\thetheorem}{\Alph{appendixc}.\arabic{theorem}}
        \renewcommand{\thedefinition}{\Alph{appendixc}.\arabic{definition}}
        \renewcommand{\thecorollary}{\Alph{appendixc}.\arabic{corollary}}
        \renewcommand{\theequation}{\Alph{appendixc}.\arabic{equation}}
        \noindent{\tenbf Appendix \theappendixc #1}\par\vspace{5pt}}
\newcommand{\subappendix}[1] {\vspace{12pt}
        \refstepcounter{subappendixc}
        \noindent{\bf Appendix \thesubappendixc. {\kern1pt \bfit #1}}
	\par\vspace{5pt}}
\newcommand{\subsubappendix}[1] {\vspace{12pt}
        \refstepcounter{subsubappendixc}
        \noindent{\rm Appendix \thesubsubappendixc. {\kern1pt \tenit #1}}
	\par\vspace{5pt}}
\newcommand{\textlineskip}{\baselineskip=13pt}
\newcommand{\smalllineskip}{\baselineskip=10pt}

\def\eightcirc{
\begin{picture}(0,0)
\put(4.4,1.8){\circle{6.5}}
\end{picture}}
\def\eightcopyright{\eightcirc\kern2.7pt\hbox{\eightrm c}} 
\newcommand{\copyrightheading}[1]
	{\vspace*{-2.5cm}\smalllineskip{\flushleft
	{\footnotesize International Journal of Modern Physics A, #1}\\
	{\footnotesize $\eightcopyright$\, World Scientific Publishing
	 Company}\\
	 }}

\newcommand{\publisher}[2]{{\begin{center}\footnotesize\smalllineskip 
	Received #1\\
	Revised #2
	\end{center}
	}}

\def\abstracts#1#2#3{{
	\centering{\begin{minipage}{4.5in}\baselineskip=10pt\footnotesize
	\parindent=0pt #1\par 
	\parindent=15pt #2\par
	\parindent=15pt #3
	\end{minipage}}\par}} 

\newcommand{\bibit}{\nineit}

\renewenvironment{thebibliography}[1]
	{\frenchspacing
	 \ninerm\baselineskip=11pt
	 \begin{list}{\arabic{enumi}.}
	{\usecounter{enumi}\setlength{\parsep}{0pt}
	 \setlength{\leftmargin 12.7pt}{\rightmargin 0pt} 
	 \setlength{\itemsep}{0pt} \settowidth
	{\labelwidth}{#1.}\sloppy}}{\end{list}}
\newcounter{itemlistc}
\newcounter{romanlistc}
\newcounter{alphlistc}
\newcounter{arabiclistc}

\newcommand{\fcaption}[1]{
        \refstepcounter{figure}
        \setbox\@tempboxa = \hbox{\footnotesize Fig.~\thefigure. #1}
        \ifdim \wd\@tempboxa > 5in
           {\begin{center}
        \parbox{5in}{\footnotesize\smalllineskip Fig.~\thefigure. #1}
            \end{center}}
        \else
             {\begin{center}
             {\footnotesize Fig.~\thefigure. #1}
              \end{center}}
        \fi}

\newcommand{\tcaption}[1]{
        \refstepcounter{table}
        \setbox\@tempboxa = \hbox{\footnotesize Table~\thetable. #1}
        \ifdim \wd\@tempboxa > 5in
           {\begin{center}
        \parbox{5in}{\footnotesize\smalllineskip Table~\thetable. #1}
            \end{center}}
        \else
             {\begin{center}
             {\footnotesize Table~\thetable. #1}
              \end{center}}
        \fi}
\def\@citex[#1]#2{\if@filesw\immediate\write\@auxout
	{\string\citation{#2}}\fi
\def\@citea{}\@cite{\@for\@citeb:=#2\do
	{\@citea\def\@citea{,}\@ifundefined
	{b@\@citeb}{{\bf ?}\@warning
	{Citation `\@citeb' on page \thepage \space undefined}}
	{\csname b@\@citeb\endcsname}}}{#1}}

\newif\if@cghi
\def\cite{\@cghitrue\@ifnextchar [{\@tempswatrue
	\@citex}{\@tempswafalse\@citex[]}}
\def\citelow{\@cghifalse\@ifnextchar [{\@tempswatrue
	\@citex}{\@tempswafalse\@citex[]}}
\def\@cite#1#2{{$\null^{#1}$\if@tempswa\typeout
	{IJCGA warning: optional citation argument 
	ignored: `#2'} \fi}}

\def\pmb#1{\setbox0=\hbox{#1}
	\kern-.025em\copy0\kern-\wd0
	\kern.05em\copy0\kern-\wd0
	\kern-.025em\raise.0433em\box0}

\def\fnt#1#2{\footnotetext{\kern-.3em
	{$^{\mbox{\scriptsize #1}}$}{#2}}}
\def\fpage#1{\begingroup
\voffset=.3in
\thispagestyle{empty}\begin{table}[b]\centerline{\footnotesize #1}
	\end{table}\endgroup}
\def\runninghead#1#2{\pagestyle{myheadings}
\markboth{{\protect\footnotesize\it{\quad #1}}\hfill}
{\hfill{\protect\footnotesize\it{#2\quad}}}}
\font\tenrm=cmr10
\font\tenit=cmti10 
\font\tenbf=cmbx10
\font\bfit=cmbxti10 at 10pt
\font\ninerm=cmr9
\font\nineit=cmti9

\font\eightrm=cmr8

\def\qed{\hbox{${\vcenter{\vbox{			
   \hrule height 0.4pt\hbox{\vrule width 0.4pt height 6pt
   \kern5pt\vrule width 0.4pt}\hrule height 0.4pt}}}$}}

\newcommand{\beq}{\begin{equation}}
\newcommand{\eeq}{\end{equation}}
\newcommand{\beqar}{\begin{eqnarray}}
\newcommand{\eeqar}{\end{eqnarray}}
\newcommand{\Pslash}{\mbox{$\not \! P$}}
\newcommand{\Eq}[1]{Eq.~(\ref{#1})}
\newcommand{\Eqs}[1]{Eqs.~(\ref{#1})}
\newcommand{\Fig}[1]{Fig.~\ref{#1}}

\begin{document}

\runninghead{D. Blaschke et al.} {Finite $T$ Meson Correlations $\ldots$}

\normalsize\textlineskip
\thispagestyle{empty}
\setcounter{page}{1}

\copyrightheading{}			

\vspace*{0.88truein}

\fpage{1}
\centerline{\bf FINITE $T$ MESON CORRELATIONS AND QUARK DECONFINEMENT
\footnote{
Supported by DAAD and NSF under Grant Nos. INT96-03385 and PHY97-22429. }}
\vspace*{0.37truein}
\centerline{\footnotesize D. BLASCHKE
\footnote{E-mail address: david.blaschke@physik.uni-rostock.de}, 
G. BURAU}
\vspace*{0.015truein}
\centerline{\footnotesize\it Fachbereich Physik, Universit\"at Rostock, 
Universit\"atsplatz 1}
\baselineskip=10pt
\centerline{\footnotesize\it D-18051 Rostock, Germany}
\vspace*{10pt}
\centerline{\footnotesize YU.L. KALINOVSKY}
\vspace*{0.015truein}
\centerline{\footnotesize\it Laboratory for Computing Techniques and 
Automation, Joint Institute for Nuclear Research}
\baselineskip=10pt
\centerline{\footnotesize\it 141980 Dubna, Russia}
\vspace*{10pt}
\centerline{\footnotesize P. MARIS, P.C. TANDY}
\vspace*{0.015truein}
\centerline{\footnotesize\it Center for Nuclear Research, Department of 
Physics, Kent State University}
\baselineskip=10pt
\centerline{\footnotesize\it Kent, OH 44242, USA}
\vspace*{0.225truein}
\publisher{(received date)}{(revised date)}

\vspace*{0.21truein}
\abstracts{
Finite temperature spatial $\bar q q$ correlation modes in the $\pi$ and
$\rho$ channels are studied with the rainbow-ladder truncated quark
Dyson-Schwinger equation and Bethe-Salpeter equation in the Matsubara
formalism.  To retain the finite range of the effective interaction
while facilitating summation over fermion Matsubara modes necessary to
ensure continuity at \mbox{$T=0$}, a separable kernel is used.  The
model is fixed by \mbox{$T=0$} properties and it implements dynamical
chiral symmetry breaking and quark confinement.    Above and below the
deconfinement and chiral restoration transitions, we study 
$M_\pi(T)$, $f_\pi(T)$ and the 3-space transverse and longitudinal 
masses $M_\rho^T(T)$ and $M_\rho^L(T)$.  For the $\rho$ mode we also 
study the strong and electromagnetic partial widths to determine
the $T$-dependence that is generated
by the quark content.  The \mbox{$\rho \to \pi \pi$} width decreases
sharply just above the chiral restoration transition leaving the
$\rho$ with only its narrow \mbox{$e^+ e^-$} width.  We discuss
improvements needed by this model, especially in regard to the  high 
$T$ behavior of the masses where we compare to lattice QCD simulations. 
}{ }{PACS numbers: 11.10.St,
11.10.Wx, 12.38.Mh, 14.40.-n, 24.85.+p}


\vspace*{1pt}\textlineskip	
\section{Introduction}		
\vspace*{-0.5pt}
\noindent
The connection between hot dense hadronic matter and a plasma of quarks
and gluons is receiving increased attention with the advent of the
relativistic heavy-ion collider (RHIC) at Brookhaven to complement
previous investigations at the CERN SPS \cite{qm99}.  The plasma is
expected to reveal itself through modified properties of hadronic
reactions and their products.  The di-lepton spectra has been of
considerable interest~\cite{RW99} as a relatively clean signal of how
vector meson correlations and their decay channels and widths might be
influenced by a hot and dense environment.  The question of how a vector
meson strong decay, such as \mbox{$\rho \to \pi \pi$}, might respond as
the temperature $T$ or chemical potential $\mu$ crosses the critical
phase boundary for chiral restoration or quark deconfinement requires
that this process be studied at the quark-gluon level.  To this end it
is desirable to be able to describe quark deconfinement and chiral
restoration at finite $T$ and $\mu$ in a manner that can be extended to
a variety of hadronic observables in a rapid and transparent way.  The
present model provides a simple framework for such investigations.
 
At \mbox{$T=\mu=0$} significant progress has been made within a
continuum approach to modeling non-perturbative QCD based on the
truncated Dyson-Schwinger equations (DSEs)~\cite{dserev,rs,AS}.  An
attractive feature of this approach is that dynamical chiral symmetry
breaking and quark confinement can be embodied in the infrared structure
of the dressed gluon 2-point function which is constrained by a few
chiral meson observables.  Recent works have employed the ladder-rainbow
truncation of the coupled quark DSE and $\bar{q}q$ Bethe-Salpeter
equation (BSE) to produce successful descriptions of masses and decay
constants of the light pseudoscalars $\pi$ and $K$~\cite{MR97} and the
vectors $\rho$, $\phi$, and $K^\star$~\cite{MT99}.  These works have
incorporated the one-loop renormalization group evolution of scale
characteristic of QCD.  With a few exceptions~\cite{MTem00},
applications to other hadron observables, including electromagnetic form
factors and coupling constants such as $g_{\rho \pi\pi}$, have usually
required the use of simpler models~\cite{T97rev,MR98,Dub98,IKR99}.

The finite $T,\mu$ extension of realistic DSE/BSE models receives extra
complications due to the breaking of $O(4)$ symmetry and the dynamical
coupling between Matsubara modes~\cite{rs,bbkr,BPRST98}. These
difficulties are compounded in the subsequent generation of hadronic
observables via straightforward adaptation of the approach~\cite{T97rev}
found to be successful at \mbox{$T=\mu=0$}.  Studies of hadronic
observables at finite $T,\mu$ in DSE/BSE models have been restricted to
simplifications such as extensions of the infrared-dominant (ID)
model~\cite{mn} in which the effective gluon propagator is restricted to
an integrable singularity at zero momentum.  Nevertheless such a model
has proved capable of yielding qualitatively useful
information~\cite{brs,mrs,dqppr}.

In this work we explore a separable Ansatz that can implement the
essential qualitative features of DSE/BSE models at finite temperature.
Separable representations have previously been found capable of an
efficient modeling of the effective $\bar q q$ interaction in the
infrared domain for \mbox{$T=0$} meson observables~\cite{sep,b+}.  A
previous implementation at \mbox{$T>0$} employed an instantaneous
separable interaction without a quark confinement
mechanism~\cite{SBK94}.  Here the approach is covariant and the simple
separable interaction absolutely confines quarks at \mbox{$T=0$}.  The
few parameters are fixed by $\pi$ and $\rho/\omega$ properties.  The
approach is a simplification of one developed earlier~\cite{b+} that was
found to be quite successful for the light meson spectrum at
\mbox{$T=0$}.  The confining mechanism is an infrared enhancement in the
quark-quark interaction that is strong enough to remove the possibility
of a mass shell pole in the quark propagator for real $p^2$.  In the
simple implementation used here, it is particularly transparent that
sufficiently high temperature will necessarily restore a quark
mass-shell pole and there will be a deconfinement transition.  This
model also implements low temperature dynamical chiral symmetry breaking
and it preserves the Goldstone theorem in that the generated $\pi$ is
massless in the chiral limit.  The solutions of the BSE for the $\pi$
and $\rho$ modes are particularly simple and are used to study the
$T$-dependence of the meson masses and decay constants in the presence
of both deconfinement and chiral restoration mechanisms.  Some
preliminary results have been previously discussed for
\mbox{$T>0 $}~\cite{bkt} and for \mbox{$T, \mu >0 $}~\cite{bt}. 

In Sec.~II the \mbox{$T=0$} separable model is introduced, the DSE
solutions for the dressed quark propagator are developed and the quark
confinement property is described.  The BSE solutions for $\pi$ and
$\rho$ are also treated there and the corresponding decay constants are
defined.  The extension to \mbox{$T>0$} is considered in Sec.~III and
the chiral symmetry restoration and deconfinement phenomena are
discussed.  The spatial $\pi$ and $\rho$ modes are considered there.
The Gell-Mann--Oakes--Renner (GMOR) relation and its generalization are
used to evaluate the performance of this model in respecting chiral
symmetry constraints.  The widths of the transverse 
$\rho$ from \mbox{$\rho \to e^+e^-$} and \mbox{$\rho \to \pi \pi $} are 
also considered in Sec.~III.   The high $T$ behavior of the model
is considered in Sec.~IV  by comparison of masses with results from
lattice simulations.  A discussion follows in Sec.~V.  An Appendix
details the high $T$ behavior of the ID model as a reference.

\section{Confining separable Dyson-Schwinger equation model}

Mesons can be described as $q \bar{q}$ bound states using the
Bethe-Salpeter equation.  In the ladder truncation, this equation
reads\footnote{We use a Euclidean space formulation, with
$\{\gamma_\mu,\gamma_\nu\}=2\delta_{\mu\nu}$, $\gamma_\mu^\dagger =
\gamma_\mu$ and $a\cdot b=\sum_{i=1}^4 a_i b_i$.}
\begin{equation}
- \lambda(P^2) \Gamma(p,P)  =  \frac{4}{3} \int \frac{d^4q}{(2\pi)^4} 
D_{\mu\nu}^{\rm eff}(p-q) \gamma_\mu S(q_+) \Gamma(q,P)S(q_-) \gamma_\nu~,
\label{bs}
\end{equation}
where $P$ is the total momentum, \mbox{$q_\pm = q \pm P/2$}, and
$D_{\mu\nu}^{\rm eff}(k)$ an ``effective gluon propagator''.  The meson
mass is identified from \mbox{$\lambda(P^2=-M^2)= 1$}.  In conjunction
with the rainbow truncation for the quark DSE
\begin{eqnarray}
\label{gendse}
 S(p)^{-1} & = & Z_2\,i\gamma\cdot p + Z_2\,m_0
   +  \frac{4}{3} \int \frac{d^4q}{(2\pi)^4} \,g^2 D_{\mu\nu}^{\rm eff}(p-q) 
                        \gamma_\mu S(q)\gamma_\nu \,.
\end{eqnarray}
this equation forms the basis for the DSE approach to meson
physics~\cite{dserev,rs}.

Recent studies have employed \mbox{$D_{\mu\nu}^{\rm eff}(k) = {\cal
G}(k^2) D_{\mu\nu}^{\rm free}(k)$} where $D_{\mu\nu}^{\rm free}(k)$ is
the free gluon propagator in Landau gauge.  The effective coupling
${\cal G}(k^2)$ is given by the one-loop perturbative form of the QCD
running coupling in the ultraviolet while the phenomenological infrared
form is chosen to reproduce the pion and kaon masses and decay
constants~\cite{MR97}.  This model can succesfully describe the vector
meson masses and dominant decays~\cite{MT99}, as well as the pion and
kaon electromagnetic form factors~\cite{MTem00}, without new parameters.

The direct extension of such an approach to accommodate finite
temperature~\cite{MRST00} and baryon density entails a significant
increase in complexity.  In the Matsubara formalism, the number of
coupled equations represented by \Eqs{bs} and (\ref{gendse}) scales up
with the number of fermion Matsubara modes included.  For studies near
and above the transition, \mbox{$T \ge 100 $}~MeV, about $10$ such modes
appear adequate~\cite{MRST00}.  The appropriate number can be $\sim
10^3$ if reasonable continuity with \mbox{$T=0$} results is to be
verified.  Meson $\bar q q$ modes at \mbox{$T > 0$} have often been
studied within the Nambu--Jona-Lasinio model where the contact nature of
the effective interaction allows decoupling of the Matsubara modes and
also analytic methods of summation, see for example, Ref.~\cite{FF94}
and references therein.  However the lack of quark confinement in that
model leads to unphysical thresholds for $\bar q q$ dissociation.

We consider here a simple separable interaction that has a finite range,
accommodates quark confinement, and facilitates a decoupling of fermion
Matsubara modes.  We base our approach on a confining separable 
model~\cite{b+} previously found to be successful at $T=0$ and defined by
\mbox{$D_{\mu\nu}^{\rm eff}(p-q) \to \delta_{\mu\nu}\, D(p^2,q^2,p \cdot q)$}
with
\begin{equation}
D(p^2,q^2,p \cdot q) = D_0~ f_0(p^2)f_0(q^2) 
                          + D_1~ f_1(p^2)(p\cdot q)f_1(q^2) \; .  
\label{model}
\end{equation}
Here a Feynman-like gauge is chosen for phenomenological simplicity.
This is a rank-2 interaction with two strength parameters $D_0, D_1$,
and corresponding form factors $f_i(p^2)$.  The choice for these
quantities is constrained by consideration of the resulting solution of
the DSE for the quark propagator in the rainbow approximation.  For the
amplitudes defined by $S(p)=[i \rlap/p A(p^2) + B(p^2) + m_0]^{-1}$ this
produces\footnote{We choose the interaction form factors such that they
provide sufficient ultraviolet suppression.  Therefore no
renormalization is needed and $Z_2=1$ .}
\begin{eqnarray}
\label{dseA}
B(p^2) &=& 
\frac{16}{3} \int \frac{d^4q}{(2\pi)^4} D(p^2,q^2,p \cdot q) 
\frac{B(q^2)+m_0}{q^2A^2(q^2)+\left[ B(q^2)+m_0\right]^2} \,\,\, , \\
\label{dseB}
\left[ A(p^2)-1 \right]  p^2 &=& 
\frac{8}{3} \int \frac{d^4q}{(2\pi)^4} D(p^2,q^2,p \cdot q) 
\frac{(p\cdot q) A(q^2)}{q^2A^2(q^2)+\left[ B(q^2)+m_0\right]^2} \,\,\, .
\end{eqnarray} 

We note that if terms of higher order in \mbox{$p \cdot q$} were to be
included in \Eq{model}, they would make no contribution to \Eqs{dseA}
and (\ref{dseB}) for the DSE solution~\cite{b+}.  The solution for
$B(p^2)$ is determined only by the $D_0$ term, and the solution for
\mbox{$A(p^2)-1$} is determined only by the $D_1$ term.  \Eq{model}
produces solutions having the form
\beq
B(p^2)=b\; f_0(p^2)\;, \;\;\;\;\; A(p^2)=1+a \;f_1(p^2)\;,
\label{ABform}
\eeq
and \Eqs{dseA} and (\ref{dseB}) reduce to nonlinear equations for the
constants $b$ and $a$. 

\subsection{Confinement and Dynamical Chiral Symmetry Breaking}

If there are no poles in the quark propagator $S(p)$ for real timelike
$p^2$ then there is no physical quark mass shell, quarks cannot
propagate free of interactions, and the description of hadronic
processes will not be hindered by spurious quark production thresholds.
This is sufficient but not necessary for quark confinement and it
remains a viable possibility for how quark confinement is
realized~\cite{RWK92,M95}.  The more general phenomena of confinement of
colored multi-quark states and the non-existence of S-matrix elements
connecting them to hadronic states are more subtle topics that do not
concern us here~\cite{dserev}.  A nontrivial solution for $B(p^2)$ in
the chiral limit ($m_0=0$) signals dynamical chiral symmetry breaking.
There is a connection between quark confinement as the lack of a quark
mass shell and the existence of a strong quark mass function in the
infrared through dynamical chiral symmetry breaking.  This connection
has proved to be empirically successful for the description of ground
state light quark mesons and their form factors and
decays~\cite{T97rev,Dub98}.  In the present separable model the strength
\mbox{$b=B(0)$}, which is generated by solution of
\Eqs{dseA} and (\ref{dseB}), controls both confinement and dynamical
chiral symmetry breaking.  

The propagator is confining if \mbox{$m^2(p^2) \neq -p^2$} for real
$p^2$ where the quark mass function is
\mbox{$m(p^2)=(B(p^2)+m_0)/A(p^2)$}.  
With an exponential form factor \mbox{$f_0(p^2)=$}
\mbox{exp $(-p^2/\Lambda_0^2)$}, this condition is most transparent in 
the case of a rank-1 separable model where $D_1=0$ and
\mbox{$A(p^2)=1$}, i.e., \mbox{$a=0$}.  In the chiral limit, the model
is confining if $D_0$ is strong enough to make
$b/\Lambda_0\ge1/\sqrt{2{\rm e}}$ and this finite $b$ also signals
dynamical chiral symmetry breaking.  Using $\Lambda_0 \sim 0.6-0.8$ GeV
as a typical range for a quark mass function $m(p^2)$, both confinement
and dynamical chiral symmetry breaking will be compatible with
\mbox{$m(p=0)\ge 0.3$} GeV, an empirically viable value.   The above
qualitative properties will also hold in the case of a rank-2 model if
the form factor ranges satisfy \mbox{$\Lambda_1 > \Lambda_0$} and this
is compatible with empirical findings that the amplitude $A(p^2)$
typically has a larger momentum range than $B(p^2)$.  At finite
temperature, the strength $b(T)$ for the quark mass function will
decrease with $T$ so that this model can be expected to have a
deconfinement transition at or before the chiral restoration transition
associated with \mbox{$b(T) \to 0$}.

It is found that the simple choice \mbox{$f_i(p^2) =$}
\mbox{exp $(-p^2/\Lambda_i^2)$} produces numerical solutions that describe
infrared properties of $\pi$ and $\omega$ mesons very well and generate
an empirically acceptable chiral condensate.   At the same time the 
produced quark propagator is found to be confining and the infrared
strength and shape of the quark amplitudes $A(p^2)$ and $B(p^2)$
are in qualitative agreement with the results of typical DSE 
studies~\cite{MR97}.    We use the exponential form factors as a minimal
way to preserve these properties while realizing that the ultraviolet 
suppression is much greater than the power law fall-off  (with
logarithmic corrections) known from asymptotic QCD.   Most of our 
investigation centers on physics below and in the vicinity of the 
transition.    We use the high $T$ behavior of masses in comparision
with lattice results to discuss improvements appropriate for future
work. 

\subsection{$\pi$ and $\rho$ bound states}

With the separable interaction of \Eq{model}, the 
allowed form of the solution of \Eq{bs} for the pion BS amplitude 
$\vec{\tau}\, \Gamma_\pi(q;P)$ is ~\cite{b+} 
\begin{equation}
\Gamma_\pi(q;P) =\gamma_5 \left(i E_\pi (P^2) +  
                     \Pslash F_\pi(P^2)\right) \; f_0(q^2) \; ,
\label{pibsa}
\end{equation} 
which contains the two dominant covariants from the set of four general
covariants.  The $q$ dependence is described only by the first form
factor $f_0(q^2)$.  The second term $f_1$ of the interaction can
contribute only indirectly via the quark propagators.  The $\pi$ BSE,
\Eq{bs}, becomes a $2\times 2$ matrix eigenvalue problem
\mbox{${\cal K}(P^2) f = \lambda(P^2) f$} where the eigenvector is
\mbox{$f = (E_\pi, F_\pi)$}.   The kernel is
\begin{equation}
{\cal K}_{ij}(P^2) = - \frac{4 D_0}{3}\, {\rm tr_s}\, \int\, 
       \frac{d^4q}{(2\pi)^4}f_0^2(q^2) 
 \left[ \hat{t}_i\, S(q_+)\, t_j\, S(q_-)\,  \right]~,
\label{pikernel}
\end{equation}
where the $\pi$ covariants are \mbox{$t=(i\gamma_5, \gamma_5\,
\Pslash)$} and we have also introduced \mbox{$\hat{t}=(i\gamma_5,$}
\mbox{$-\gamma_5\, \Pslash/2P^2)$}. We note that the separable model
produces the same $q^2$ shape for both amplitudes $F_\pi$ and $E_\pi$;
the shape is that of the quark amplitude $B(q^2)$.  For the general
amplitude $E_\pi(q;P)$ this is the correct shape in the chiral limit;
for physical quark masses it is still a very good
approximation~\cite{MR97}.  In general, the amplitude $F_\pi(q;P)$ does
not have that shape; it is in fact linked with $A(q^2)-1$ through the
axial vector Ward-Takahashi identity (AV-WTI)~\cite{MRT98}.  However
Goldstone's theorem, the key consequence of the AV-WTI, is preserved by
the present separable model; in the chiral limit, whenever a nontrivial
DSE solution for $B(p^2)$ exists, there will be a massless $\pi$
solution to \Eq{pikernel}. 

A simple illustrative truncation is obtained by setting \mbox{$F_\pi(P^2)=0$}
for then \Eq{pikernel} reduces to an expression for the eigenvalue which is
\begin{equation}
\lambda_\pi(P^2) = 
\frac{16 D_0}{3} \int \frac{d^4q}{(2\pi)^4}f_0^2(q^2) 
\left[\left(q^2-\frac{P^2}{4}\right) \sigma_V^+ \sigma_V^- 
                                       + \sigma_S^+ \sigma_S^- \right]~,
\label{lampi}
\end{equation}
where the quark propagator amplitudes employed here are defined by 
\mbox{$S(p)=$} \mbox{$-i\rlap/p$} \mbox{$\sigma_V(p^2) + $} 
\mbox{$\sigma_S(p^2)$}.    Since \mbox{$A=1$} in rank-1,  we  omit the
amplitude $F_\pi$ in that case.   Thus it is \Eq{lampi} that we use for the
$\pi$ calculation with a rank-1 interaction.

The vector mesons $\rho$ and $\omega$ are degenerate in the ladder
approximation.   We shall deal with the isovector $\rho$.   For the 
BS amplitude $\vec{\tau}\,\Gamma^\rho_\mu(q;P)$ there are in general eight 
transverse covariants~\cite{MT99} and the dominant one is 
\mbox{$\gamma_\mu^T(P)=$} \mbox{$T_{\mu \nu}(P)\gamma_\nu$} where 
\mbox{$T_{\mu \nu}(P)=\delta_{\mu \nu} - P_\mu P_\nu /P^2$}.
The solution of the rank-1 model contains only that term, that is,
\mbox{$\Gamma^\rho_\mu(q;P) =$}
\mbox{$\gamma_\mu^T(P) f_0(q^2) F_\rho(P^2)$}.   The corresponding
eigenvalue is given by
\begin{equation}
\lambda_\rho(P^2) = 
\frac{8 D_0}{3} \int \frac{d^4q}{(2\pi)^4}f_0^2(q^2) 
\left[\left(q^2-\frac{P^2}{4}-\frac{2q^2}{3}(1-z^2) \right) \sigma_V^+ 
                              \sigma_V^-  + \sigma_S^+ \sigma_S^- \right]~,
\label{lamv}
\end{equation}
where \mbox{$z=\hat{q} \cdot \hat{P}$}.  For the rank-2 separable
interaction, there are two other covariants besides $\gamma_\mu^T$ that
will appear in the solution of the BSE~\cite{b+}.  However it has been
found in such a model that these subleading vector covariants make only
a few percent contribution to the vector mass~\cite{b+} and the
associated $g_{\rho\pi\pi}$~\cite{Dub98}.  In this present work we will
ignore the subdominant covariants and employ $\gamma_\mu^T$ as the only
covariant for the vector meson.  The differences in the vector mass
obtained from Eq.~(\ref{lamv}) in rank-1 and rank-2 will be due to
differences in the quark propagator in each case.

The normalization condition for the $\pi$ BS amplitude can be expressed as
\begin{eqnarray}
\label{pinorm}
    \left. 2 P_\mu = \frac{\partial}{\partial P_\mu} \,
       2 N_c \, {\rm tr}_s \int \frac{d^4q}{(2\pi)^4} 
        \bar\Gamma_\pi(q;-K)\, S(q_+)\, 
        \Gamma_\pi(q;K)\,S(q_-)
        \right|_{P^2=K^2=-M_\pi^2} \,.
\end{eqnarray}
Here $\bar\Gamma(q;K)$ is the charge conjugate amplitude \mbox{$[{\cal
C}^{-1} \Gamma(-q,K) {\cal C}]^{\rm t}$} where \mbox{${\cal C}=\gamma_2
\gamma_4$} and t denotes matrix transpose.  This defines a normalization
constant $N_\pi$ via \mbox{$E_\pi(P^2=-M_\pi^2)\, f_0(q^2)=$}
\mbox{$B(q^2)/N_\pi$}.  The pion decay constant $f_\pi$ can be expressed
as the loop integral
\begin{eqnarray}
   f_\pi \; P_\mu\,\delta_{ij} &=& \langle 0|\bar q \frac{\tau_i}{2}
   \gamma_\mu \gamma_5 q | \pi_j(P)\rangle
\nonumber \\
        &=& \delta_{ij} \, N_c \, {\rm tr_s} \int \frac{d^4q}{(2\pi)^4}\;
     \gamma_5 \gamma_\mu \; S(q_+) \; \Gamma_\pi(q;P)\; S(q_-)\; .
\label{fpi}
\end{eqnarray}
Note that \Eq{fpi} is the exact expression for $f_\pi$ except for
the absence of the renormalization constant $Z_2$ which is unity in the
present model.  The decay constants thus test the quality of the
infrared behavior of the quark propagator and the BS amplitudes.
Similar expressions exist for the normalization of the vector meson BS
amplitude and the coupling between a vector meson and a
photon~\cite{MT99,IKR99}.
\noindent
\begin{table}[ht]
\tcaption{\label{results}
Calculated properties at \mbox{$T=0$} for the $u/d$ quark  mesons $\pi$ 
and $\rho$ along with related quark properties and parameters for the
rank-1 and rank-2 separable models.  The quoted experimental values for both
the quark condensate and the current $u/d$ quark mass  $m_0$ are appropriate
to a renormalization scale of $\mu \sim 1$ GeV.  The scale for the 
model calculations may be considered to be set by $\Lambda_0$. }  
\centerline{\footnotesize\smalllineskip
\begin{tabular}{| l | c c c |}
\hline
        & Experiment  & rank-1     & rank-2          \\ \hline
-~$\langle \bar q q \rangle^0$
                & $(0.236\,{\rm GeV})^3$ & 0.248 & 0.203  \\
$m_0$           & 5 - 10 MeV  & 6.6  &  5.3 \\
$m(p^2=0)$      & $\sim$0.350 GeV & 0.685  & 0.405 \\
$M_\pi$         &  0.1385 GeV & 0.140 & 0.139   \\
$f_\pi$         &  0.093 GeV & 0.104  & 0.093   \\
$N_\pi/f_\pi$   &  1.0       &  0.987 & 0.740   \\
$M_{\rho/\omega}$      &  0.770/0.783  GeV & 0.783 & 0.784    \\
$g_\rho$        &  5.04 & 5.04 & 6.38    \\
\mbox{$\Gamma_{\rho^0 \to e^+e^-}$} &   6.77 keV & 6.76  & 4.22  \\
$g_{\rho\pi\pi}$ &  6.05       & 5.71  & 7.22     \\ 
\mbox{$\Gamma_{\rho \to \pi\pi}$} & 151 MeV  &  137  &  221  \\   \hline
Parameters      &             &       &           \\ \hline
$D_0\Lambda_0^2$ &            & 128.0 & 260.0      \\
$\Lambda_0$     &     & 0.687 GeV     & 0.638 GeV   \\
$D_1\Lambda_1^4$ &    & 0     & 130.0      \\
$\Lambda_1/\Lambda_0$ & &      & 1.21     \\  \hline
\end{tabular}}
\end{table}
 
In Table~\ref{results} the results for the ground state $\pi$ and
$\rho/\omega$ mesons as well as related quantities are shown for both
rank-1 and rank-2 versions of the model along with the values of the
employed parameters.  We consider the experimental $\omega$ mass to be
the appropriate value for comparison with the vector result in ladder
approximation in order to allow for subsequent preferential lowering of
$M_\rho$ due to pion loop dressing~\cite{rhomeslp}.  The range parameter
$\Lambda_0$ is used to set the mass scale to reproduce the experimental
$M_\omega$ for rank-1 and $f_\pi$ for rank-2.  The obtained dynamical
quark mass function at \mbox{$p^2=0$} is
\mbox{$m(p^2=0)=0.685$~GeV} for rank-1.    For rank-2 we obtain 
\mbox{$A(p^2=0)= 1.94$} and \mbox{$m(p^2=0)= 0.405$~GeV}.  In both cases 
the dressed quark propagator is confining.   The strengths obtained
for the mass function are consistent with results from recent
DSE solutions~\cite{MR97,MT99}.  
Also shown in Table~\ref{results} are the results obtained for the
electromagnetic ($g_\rho$) and strong decays ($g_{\rho\pi\pi}$) of the 
$\rho$, which compare reasonably well with experiments.  

The difference between $N_\pi$ and $f_\pi$ shown in Table~\ref{results}
simply reflects the fact that the AV-WTI cannot be exactly satisfied within a
separable model.  (This is also the case in the Nambu--Jona-Lasinio model
due to the required cut-off and in any approach that does not use a 
translationally invariant interaction or a sufficiently complete set of 
Dirac covariants for $\Gamma_\pi$ and the vertex amplitudes~\cite{MRT98}.) 
A translationally invariant model that makes the simplifying 
assumption \mbox{$A(p^2)=1$} can achieve a fortuitious 
agreement~\cite{dserev} and this dominates the rank-1 result for 
\mbox{$N_\pi/f_\pi$}.  Rather than use $f_\pi$ to define $N_\pi$, we 
calculate $N_\pi$ from its definition in \Eq{pinorm} so that, in subsequent 
studies of processes such as \mbox{$\rho \to \pi \pi$}, the pion state is 
physically and consistently normed within the model.

\section{Finite Temperature Extension}
\label{sec:mesons}

The extension of the separable model studies  to 
$T\neq 0$ is systematically accomplished by 
transcription of the Euclidean quark 4-momentum via \mbox{$q \rightarrow$}
\mbox{$ q_n =$} \mbox{$(\omega_n, \vec{q})$}, where 
\mbox{$\omega_n=(2n+1)\pi T$} are the discrete Matsubara frequencies. 
The effective $\bar q q$ interaction will automatically decrease with 
increasing $T$ without the introduction of an explicit $T$-dependence
which would require new parameters.   We investigate the resulting 
behavior of the $\pi$ and $\rho$ meson modes and decays in the presence of
deconfinement and chiral restoration.   

\subsection{Chiral symmetry restoration and deconfinement}

\begin{figure}[ht]
\centerline{
\epsfig{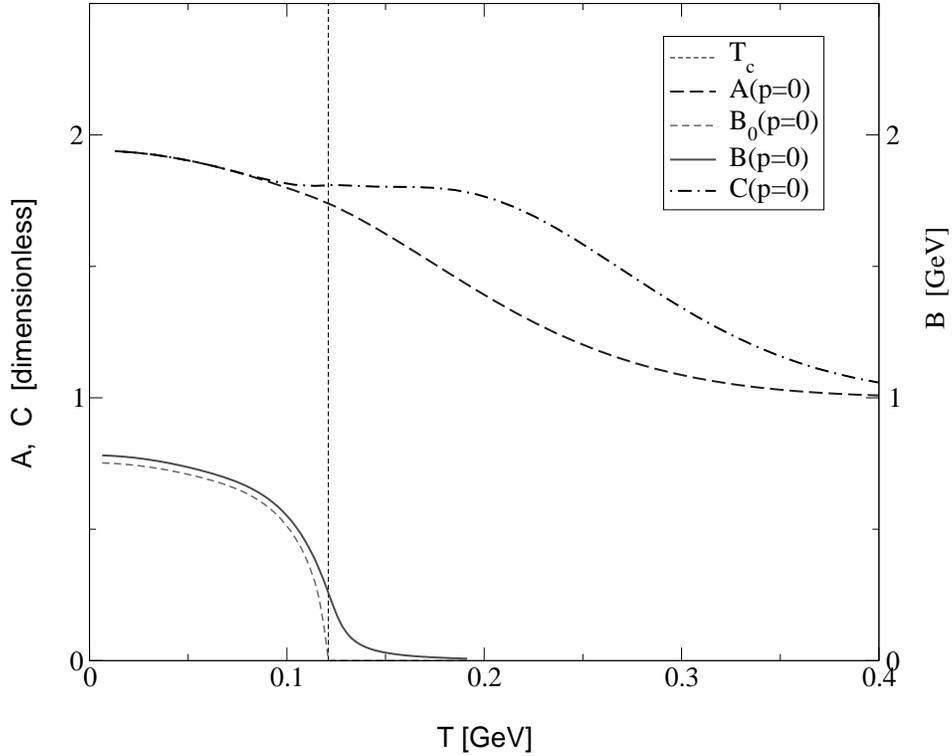}
}
\caption{\label{fig:ABC} $T$-dependence of quark self-energy amplitudes 
at \mbox{$p=0$} from solution of the DSE. } 
\end{figure}
The result of the DSE solution for the dressed quark propagator now becomes
\beq
S^{-1}(p_n, T) = i\vec{\gamma} \cdot \vec{p}\; A(p_n^2,T)
                            + i \gamma_4 \omega_n\; C(p_n^2,T)
                            + B(p_n^2,T) + m_0 \; ,
\label{invprop}
\eeq
where \mbox{$p_n^2=\omega_n^2 + \vec{p}^{\,2}$} and there are now three
amplitudes due to the loss of $O(4)$ symmetry.  The solutions have the
form \mbox{$B= b(T) f_0(p_n^2)$}, \mbox{$A=1+ a(T) f_1(p_n^2)$}, and
\mbox{$C=1+ c(T) f_1(p_n^2)$} and the DSE becomes a set of three
non-linear equations for $ b(T)$, $a(T)$ and $c(T)$.  The explicit form
is
\begin{equation}
 a(T) = \frac{8 D_1}{9}\,  T \sum_n \int \frac{d^3p}{(2\pi)^3}\,
 f_1(p_n^2)\, \vec{p}^{\,2}\, [1 +  a(T) f_1(p_n^2)]\; d^{-1}(p_n^2,T) \; ,
\end{equation}

\begin{equation}
 c(T) = \frac{8 D_1}{3}\,  T \sum_n \int \frac{d^3p}{(2\pi)^3}\,
 f_1(p_n^2)\, \omega_n^2\, [1 +  c(T) f_1(p_n^2)]\; 
                                                d^{-1}(p_n^2,T) \; ,
\end{equation}

\begin{equation}
 b(T) = \frac{16 D_0}{3}\,  T \sum_n \int \frac{d^3p}{(2\pi)^3}\,
 f_0(p_n^2)\, [m_0 +  b(T) f_0(p_n^2)]\; d^{-1}(p_n^2,T) \; ,
\end{equation}
where $d(p_n^2,T)$ is given by
\begin{equation}
d(p_n^2,T) = \vec{p}^{\,2}\, A^2(p_n^2,T) +\omega_n^2\,  C^2(p_n^2,T)
                  + [m_0 +  B(p_n^2,T)]^2 \; .
\end{equation}
It is also useful to introduce the equivalent representation
\beq
S(p_n, T) = -i\vec{\gamma} \cdot \vec{p}\; \sigma_A(p_n^2,T)
               -i\gamma_4 \omega_n\; \sigma_C(p_n^2,T)
               + \sigma_B(p_n^2,T) \; , 
\label{prop}
\eeq
where \mbox{$\sigma_A = A/d$}, \mbox{$\sigma_C = C/d$} and
\mbox{$\sigma_B = (B+m_0)/d$}.  

The $T$-dependence of the solutions for $A, B$ and $C$ at
\mbox{$\vec{p}^2=0$} for the rank-2 model is displayed in
Fig.~\ref{fig:ABC}.  The results shown are for the lowest Matsubara mode
($n=0$) which provides the leading behavior as $T$ is increased.  The
chiral restoration critical temperature $T_c$ is identified from the
vanishing of the chiral limit amplitude $B_0(p=0,T)$ as shown. We find
\mbox{$T_c=$} 121~MeV for the rank-2 model.  Below $T_c$, $A$ and $C$
are relatively constant, $O(4)$ symmetry is approximately manifest, and
the main effect is an almost constant quark wave function
renormalization via $1/C$; the central feature shared by both rank-1 and
rank-2 models is a rapidly decreasing mass function.  Above $T_c$, there
remains a significant temperature range where the self-interaction
effects are strong, both $A$ and $C$ are considerably enhanced above
their perturbative values, and the breaking of $O(4)$ symmetry is
manifest.  The present model thus captures the qualitative
$T$-dependence observed for the dressed quark propagator in studies of
the quark DSE~\cite{bbkr,brs}.  The rank-1 limit has \mbox{$A=C=1$} for
all $T$ and the behavior of $B(p=0,T)$ is similar to that in
\Fig{fig:ABC} except \mbox{$T_c=$} 146~MeV is obtained.

The order parameter for chiral restoration, the chiral quark condensate,
can be obtained from the chiral limit quark propagator as
\mbox{$\langle \bar q q \rangle^0 = - N_c\, {\rm tr_s} \,S_0(x,x)$}.  
Here this produces
\begin{equation}
\langle \bar q q \rangle^0 = - 4 N_c\,  T \sum_n \int \frac{d^3p}{(2\pi)^3}\,
    \frac{ b_0(T)\, f_0(p_n^2)}{d_0(p_n^2,T)} \; ,
\end{equation}
where $ b_0$ and $d_0$ are obtained from the chiral limit solution of
the DSE.  Both $ b_0(T)$ and $\langle \bar{q}q \rangle^0$ vanish sharply
as \mbox{$(1-T/T_c)^\beta$} with the critical exponent having the mean
field value \mbox{$\beta=1/2$} in agreement with other rainbow DSE
studies~\cite{BHRS98+HMR99}.

The deconfinement temperature $T_d$ is found by a search for a
propagator pole (a zero of the function $d(p_0^2,T)$) on the real
$\vec{p}^{\,2}$ axis.  We find \mbox{$T_d=T_c=$} 146~MeV for the rank-1
model and \mbox{$T_d=0.9~T_c=$} 105~MeV for the rank-2 model.  Only
about 15\% variation in these transition temperatures can be achieved by
variation of the model parameters while retaining a reasonable
description of the observables shown in Table~\ref{results}.  Since all
dynamical information concerning the rank-1 model is contained in one
function $B(p_n^2,T)$, one may expect to find \mbox{$T_d=T_c$} in that
case.  For rank-2, the dynamical information is contained in three
functions and the result \mbox{$T_d \neq T_c$} is not surprising.  It is
however possible that the separable form of interaction used here might
miss some dynamical correlations between $A, B$ and $C$ that would
otherwise produce \mbox{$T_d =T_c$}.

\subsection{Spatial $\pi$ correlations at $T\neq0$}

At \mbox{$T=0$} the mass-shell condition for a meson as a $\bar q q$
bound state of the BSE is equivalent to the appearance of a pole in the
$\bar q q$ scattering amplitude as a function of $P^2$.  At $T\neq0$ in
the Matsubara formalism, the $O(4)$ symmetry is broken by the heat bath
and we have \mbox{$P \to (\Omega_m,\vec{P})$} where \mbox{$\Omega_m = 2m
\pi T$}.  Bound states and the poles they generate in propagators may be
investigated through polarization tensors, correlators or Bethe-Salpeter
eigenvalues.  This pole structure is characterized by information at
discrete points $\Omega_m$ on the imaginary energy axis and at a
continuum of 3-momenta.  Analytic continuation for construction of
real-time Green's functions (and related propagation properties) has
been well-studied~\cite{LvW87}.  An unambiguous result is obtained by
requiring that the continuation yield a function that is bounded at
complex infinity and analytic off the real axis~\cite{LvW87}.  One may
search for poles as a function of $\vec{P}^2$ thus identifying the
so-called spatial or screening masses for each Matsubara mode.  These
serve as one particular characterization of the propagator and the
\mbox{$T > 0 $} bound states.

In the present context the eigenvalues of the meson BSE become
\mbox{$\lambda(P^2) \to $} \mbox{$\tilde{\lambda}(\Omega_m^2,\vec{P}^2;T)$}. 
The temporal meson masses identified by zeros of
$1-\tilde{\lambda}(\Omega^2,0;T)$ will be different in general from the
spatial masses identified by zeros of
$1-\tilde{\lambda}(0,\vec{P}^2;T)$.  They are however identical at
\mbox{$T =0$} and an approximate degeneracy can be expected to extend
over the finite $T$ domain where the $O(4)$ symmetry is not strongly
broken.  From \Fig{fig:ABC} one may reasonably expect this domain to
extend up to \mbox{$T \sim 80-100$}~MeV.  At and above the transition,
temporal and spatial masses can be expected to emphasize different
aspects of the bound state modes.  In this work we explore the
$T$-dependence of the lowest spatial masses in the $\pi$ and $\rho$
channels.

In the $\pi$ channel, the correlator 
\mbox{$\Pi(x) = \langle T\; J_{ps}(x)\; J_{ps}(0)\rangle$} of two currents
\mbox{$J_{ps}(x)=$} \mbox{$ \bar{q}(x)\gamma_5 q(x)$}, after transformation
to momentum space and extension to \mbox{$T >0$} via the Matsubara
formalism, can be expressed as
\begin{equation}
\Pi(\Omega^2, \vec{P}^2) =  
      N_c \; T \sum_n {\rm tr}_s \int \frac{d^3q}{(2\pi)^3}\, 
     \gamma_5\, S(q_{n+})\, \Gamma_{ps}(q_n;\Omega,\vec{P})\,  S(q_{n-})\,,
\label{Pi}
\end{equation}
where \mbox{$q_{n\pm}= q_n \pm P/2$} and $\Gamma_{ps}$ is the pseudoscalar 
vertex which satisfies the inhomogeneous version of the pion BSE.  The 
quark propagators 
appearing here are dressed according to solution of the DSE. When the ladder 
approximation is employed for $\Gamma_{ps}$, along with the present separable
approximation to the BSE kernel, the correlation function can be expressed
as 
\begin{equation}
\Pi(\Omega^2, \vec{P}^2) = \Pi^{(0)}(\Omega^2, \vec{P}^2) 
       -\frac{4D_0}{3} N_c\,
  L_i(\Omega^2, \vec{P}^2)\;[1 -{\cal K}(\Omega^2, \vec{P}^2)]^{-1}_{ij} 
                                         \; \hat{L}_j(\Omega^2, \vec{P}^2) ~.
\label{Pisep}
\end{equation}
Here $\Pi^{(0)}$ is the contribution to Eq.~(\ref{Pi}) arising from the 
zeroth order contribution ($\gamma_5$) to the pseudoscalar vertex 
$\Gamma_{ps}$.  The second term of Eq.~(\ref{Pisep}) sums the interaction
terms and factorizes due the separability of the effective interaction.
The kernel ${\cal K}$ involves the \mbox{$T>0$} extension of the kernel 
of the $\pi$ separable BSE given previously in Eq.~(\ref{pikernel}).
The loop integrals for the numerator are given by 
\begin{equation}
L_i(\Omega^2, \vec{P}^2) = T \sum_n {\rm tr}_s \int \frac{d^3q}{(2\pi)^3}\, 
          f_0(q_n^2)\, \gamma_5\, S(q_{n+})\, t_i \, S(q_{n-})\,,
\label{Li}
\end{equation}
and
\begin{equation}
\hat{L}_j(\Omega^2, \vec{P}^2) = 
                 T \sum_n {\rm tr}_s \int \frac{d^3q}{(2\pi)^3}\, 
          f_0(q_n^2)\,  \hat{t}_j \, S(q_{n+})\, \gamma_5\, S(q_{n-})\,.
\label{Lhatj}
\end{equation}

With an eigenvector representation of the BSE kernel ${\cal K}$, 
Eq.~(\ref{Pisep}) develops a denominator 
\mbox{$1-\tilde{\lambda}_\pi(\Omega^2, \vec{P}^2;T)$}.   There is a pole 
in the correlator associated with the  spatial mode solution to the 
homogeneous BSE identified from 
\begin{equation}
1-\tilde{\lambda}_\pi(0, \vec{P}^2;T) = 
                     Z_\pi^{-1}(\vec{P}^2,T) [\vec{P}^2 +M_\pi^2(T)] =0 \; .
\label{dennom}
\end{equation}
The masses so identified are spatial screening masses of the lowest mode
associated with the 3-space asymptotic behavior
\mbox{$\Pi(x) \sim \exp(-M x)$}.   The identification of spatial masses
by location of a pole is equivalent, by Fourier transformation, to the 
method of large spatial separation of sources used in lattice QCD. 

The general form of the finite $T$ pion BS amplitude allowed by the 
separable model is 
\begin{equation}
\Gamma_\pi(q_n;P_m) =\gamma_5 \left(i E_\pi (P_m^2)   
  +  \gamma_4 \, \Omega_m \tilde{F}_\pi(P_m^2)
  +  \vec{\gamma} \cdot \vec{P} F_\pi(P_m^2)\right) \; f_0(q_n^2) \; .
\label{pibsaT}
\end{equation}
The separable BSE becomes a $3\times 3$ matrix eigenvalue problem with a
kernel that is a generalization of Eq.~(\ref{pikernel}).  In the limit
\mbox{$\Omega_m \to 0$}, as is required for the spatial mode of interest
here, the amplitude \mbox{$\hat{F}_\pi = \Omega_m \tilde{F}_\pi$} is
trivially zero.  The pseudovector amplitude $F_\pi$ is significantly
different from zero below $T_c$, but decreases rapidly above the
transition.  In the chiral limit, it vanishes identically at and beyond
the transition; above the transition, only $E_\pi$ is survives.  

\begin{figure}[ht]
\centerline{
\epsfig{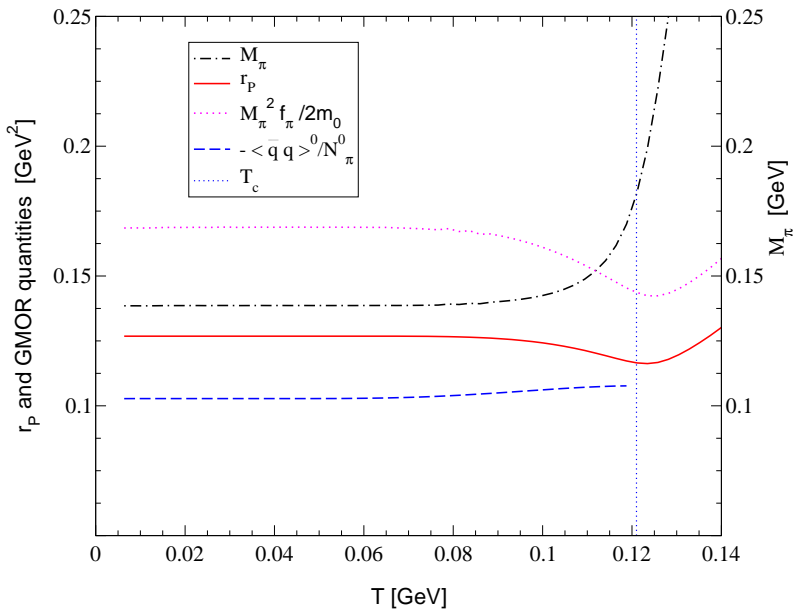}
}
\caption{\label{fig:r2pi} Chiral symmetry in the rank-2 model.  
$T$-dependence of $M_\pi$ and quantities involved in 
the exact pion mass relation and the GMOR relation.   }
\end{figure}

The result for the $\pi$ mass is displayed in Fig.~\ref{fig:r2pi} for
rank-2; the results for the rank-1 model are similar.  In both models,
$M_\pi(T)$ is seen to be only weakly $T$-dependent until near $T_c$
where a sharp rise begins.  These qualitative features of the response
of the pion mode with $T$ agree with the results deduced from the DSE in
Ref.~\cite{bbkr} and also with the more detailed study of
Ref.~\cite{MRST00} using a ladder-rainbow truncation of the DSE/BSE
system that preserves the one-loop renormalization group properties of
QCD.  Evidently the detailed character of the effective interaction in
the perturbative region does not dominate at the level of the present
qualitative investigation.

\subsection{Chiral symmetry and $\pi$ mass relation}

To explore the extent to which the model respects the detailed constraints 
from chiral symmetry, we investigate the exact QCD  pseudoscalar mass 
relation~\cite{MRT98} which, after extension to the spatial mode at 
\mbox{$T>0$}, is
\begin{equation}
\label{gen-GMOR}
M_\pi^2(T) \, f_\pi(T) = 2 m_0 \, r_P(T)~.
\end{equation}
Here $r_P$, the residue at the pion pole in the pseudoscalar 
vertex, is given by the pseudoscalar projection of the pion wavefunction 
onto zero quark-antiquark separation, that is
\begin{equation}
\label{rp}
i r_P(T) =  N_c \; T \sum_n {\rm tr}_s \int \frac{d^3q}{(2\pi)^3}\, 
     \gamma_5 S(q_n+\frac{\vec{P}}{2}) \Gamma_\pi (q_n;\vec{P}) 
                                   S(q_n-\frac{\vec{P}}{2})\,.
\end{equation}
The generalization of Eq.~(\ref{fpi}) for $f_\pi$ to finite temperature
in the case of the spatial pion mode is
\begin{equation}
\label{fpiT}
P_i\, f_\pi(T) =  N_c \; T \sum_n {\rm tr}_s \int \frac{d^3q}{(2\pi)^3}\, 
     \gamma_5 \gamma_i \; S(q_n+\frac{\vec{P}}{2}) \;\Gamma_\pi (q_n;\vec{P}) 
                               \;    S(q_n-\frac{\vec{P}}{2})\,.
\end{equation}
\Eqs{rp} and (\ref{fpiT}) are exact expressions for $r_P(T)$ and $f_\pi(T)$  
except for the absence of the renormalization constants which are
trivially equal to one in this separable model.  The relation in
Eq.~(\ref{gen-GMOR}) is a consequence of the pion pole structure of the
isovector axial Ward identity which links the quark propagator, the
pseudoscalar vertex and the axial vector vertex~\cite{MRT98}.  In the
chiral limit, \mbox{$r_P \rightarrow$}
\mbox{$- \langle \bar{q} q\rangle^0/f_\pi^0$} and Eq.(\ref{gen-GMOR}), 
for small mass, produces the Gell-Mann--Oakes--Renner (GMOR) relation.
The exact mass relation, Eq.~(\ref{gen-GMOR}), can only be satisfied
approximately when the various quantities are obtained in a manner that
does not preserve axial Ward identity.  The error can be used to assess
the reliability of the present approach to modeling the behavior of the
pion spatial mode as the temperature is varied.

Our findings in the case of the rank-2 model are displayed in
Fig.~\ref{fig:r2pi}.  There the solid line represents $r_P(T)$
calculated from the quark loop integral in Eq.~(\ref{rp}); the dotted
line represents $r_P$ constructed from the other quantities in
Eq.~(\ref{gen-GMOR}).  The exact mass relation, \Eq{gen-GMOR}, is
violated by about 25\% almost independent of temperature, even above
$T_c$.  The rank-1 model satisfies this relation to within 1\%, due to
the special but unrealistic case $A = 1 = C$.  
We have also investigated the (approximate) GMOR relation within the
present model.  The quantity \mbox{$ -\langle \bar{q}
q\rangle^0/N_\pi^0$}, displayed in Fig.~\ref{fig:r2pi}, is the chiral
limit of $r_P$ in this model.  If all covariants for the pion were
retained and the axial vector Ward identity were obeyed, one would have
\mbox{$N_\pi^0=f_\pi^0$} in the chiral limit~\cite{MRT98}.   If the GMOR 
relation were exactly obeyed, the long-dashed line representing
\mbox{$ \langle \bar{q} q\rangle^0/N_\pi^0$} would coincide with the dotted
line representing \mbox{$M_\pi^2\,f_\pi/2m_0$}.  These features are 
temperature-independent until about $0.9~T_c$, consistent with an earlier 
study of low energy theorems at \mbox{$T>0$} within a three-space, 
non-confining separable interaction model~\cite{SBK95}.  Close to $T_c$, 
the GMOR relation breaks down, and above $T_c$, it is no longer well-defined.

It should be noted that $f_\pi^0, N_\pi^0$ and $ \langle \bar{q}
q\rangle^0$ are equivalent order parameters for the critical behavior
near $T_c$ and have weak $T$-dependence below $T_c$.  A consequence is
that $M_\pi^2 \, f_\pi$, $r_P$ and 
\mbox{$ \langle \bar{q} q\rangle^0/N_\pi^0$} are almost $T$-independent 
and so are the estimated errors for the two mass relations linking these
quantities.  We have employed the physical $f_\pi$ defined at non-zero
current quark mass, and this does not vanish at $T_c$ but continuously
decreases.

\subsection{Spatial $\rho$ correlations at $T\neq0$}

\begin{figure}[ht]
\centerline{
\epsfig{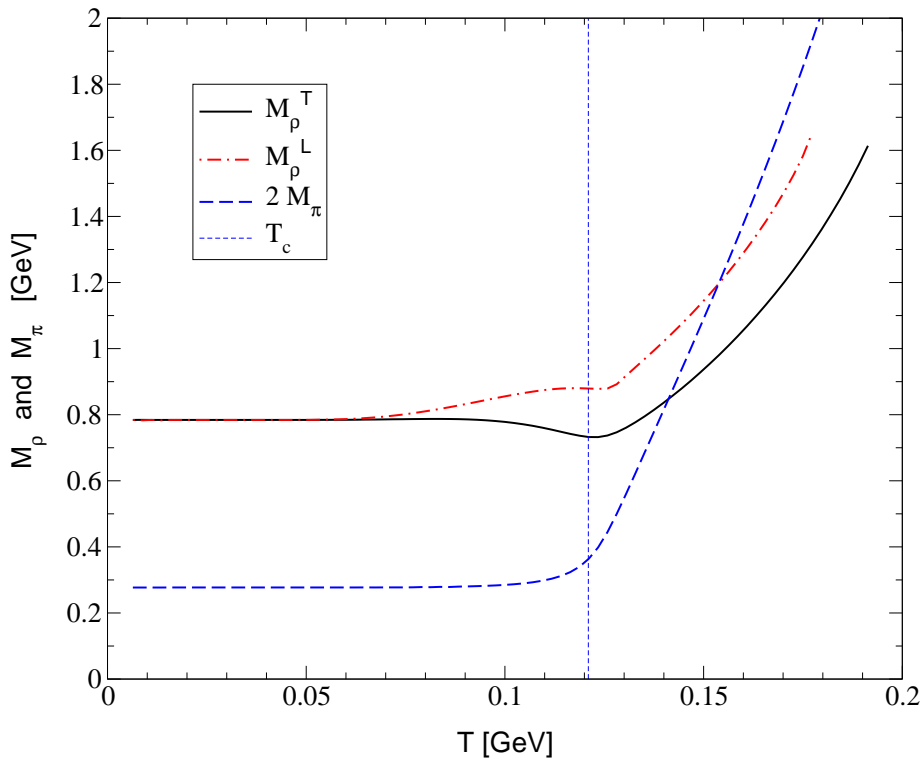}
}
\caption{\label{fig:rhomass} $T$-dependence of the longitudinal $\rho$ 
mass (dot-dashed line) and the transverse $\rho$ mass (solid line).   
Also shown is $2 M_\pi$ indicating that the strong decay channel becomes
inaccessible within 20-30 MeV beyond $T_c$.   Results are from the 
rank-2 model. }
\end{figure}

The \mbox{$T=0$} transverse vector meson, that we have described by the
covariant $\gamma_\mu^T$, splits for \mbox{$T>0$} into 3-space
longitudinal and transverse modes.  For the spatial modes characterized
by \mbox{$P=(0,\vec{P})$} the BS amplitudes are
\beq
\Gamma^{\rho(L)}_\mu(q_n;\vec{P}) = \delta_{\mu 4} \, \gamma_4
                               f_0(q_n^2) F_{\rho(L)}(\vec{P}^2)  \; ,
\label{BSrhoTL}
\eeq
and
\beq
\Gamma^{\rho(T)}_i(q_n;\vec{P}) =  
\left( \gamma_i - \frac{P_i \vec{P} \cdot \vec{\gamma} }{\vec{P}^2} \right)
                               f_0(q_n^2) F_{\rho(T)}(\vec{P}^2)  \; .
\label{BSrhoTT}
\eeq
The $T$-dependence of the corresponding masses is displayed in
Fig.~\ref{fig:rhomass} for the rank-2 model.  These modes are
effectively degenerate and $T$-independent until about $T_c/2$ where the
breaking of $O(4)$ invariance becomes significant.  The qualitative
features \mbox{$M^L_\rho (T) > M^T_\rho (T)$} and
\mbox{$ M^T_\rho (T) \approx {\rm const}$} for \mbox{$T<T_c$} seen here 
in the present context of a finite range interaction have previously
been noted within the limiting case of the zero momentum range ID
model~\cite{mrs}.  This latter model was not applied for $T>T_c$.  We
discontinue the present study of the longitudinal mode at $T \sim $ 180
MeV where it becomes unstable to $\bar q q$ dissociation.  The
transverse mode continues to be below the spatial $\bar q q$ threshold
for the temperature range displayed.

\begin{figure}[ht]
\centerline{
\epsfig{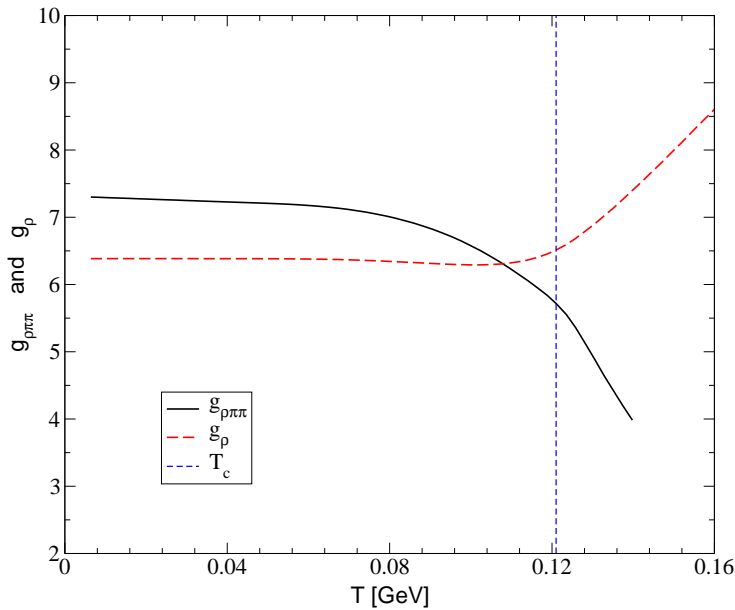}
}
\caption{\label{fig:grpp} $T$-dependence of the $\rho$ electromagnetic
decay constant $g_\rho$ (dashed line) and the strong coupling 
constant $g_{\rho \pi\pi}$ (solid line). Results are from the 
rank-2 model.   }
\end{figure}

The \mbox{$T=0$} expression~\cite{MT99,IKR99} for the electromagnetic 
coupling constant
$g_\rho$ has a straightforward extension to \mbox{$T>0$} for a
transverse spatial $\rho^0$ mode.  Use of the \mbox{$\Omega_m=0$}
solution described above yields
\begin{eqnarray}
\frac{{M^T_\rho}(T)^2}{g_\rho(T)}
 &=& \frac{N_c \, T}{2} \sum_n {\rm tr}_s \int\!\frac{d^3q}{(2\pi)^3}\,
        \gamma_i \, S(q_n+\frac{\vec{P}}{2})\, 
        \Gamma_i^{\rho(T)}(q_n;\vec{P}) \, S(q_n-\frac{\vec{P}}{2})\,.
\label{rhophoton}
\end{eqnarray}
With summation over enough Matsubara modes for convergence, the produced
$g_\rho(T)$ tends smoothly to the previously determined \mbox{$T=0$}
result.  The result over a temperature range that extends just beyond
$T_c$ is displayed in Fig.~\ref{fig:grpp}.  There it is confirmed that
there is very little $T$-dependence below about $0.9~T_c$ where there is
approximate $O(4)$ symmetry as evident in the quark propagator behavior
in Fig.~\ref{fig:ABC}.

The impulse approximation for the $\rho\pi\pi$
vertex~\cite{T97rev,Dub98}, after extension to \mbox{$T>0$} for spatial
modes characterized by \mbox{$Q=(0,\vec{Q})$} for the $\rho$ and
\mbox{$P=(0,\vec{P})$} for the relative $\pi\pi$ momentum, takes the
form
\begin{eqnarray}
\label{gpp}
\Lambda_\nu(P,Q)&=&P_\nu \, g_{\rho \pi\pi}(T) \nonumber\\
&=& -2 N_c\, T \sum_n {\rm tr}_s \int \frac{d^3q}{(2\pi)^3} \,
 \Gamma_\pi(k_{n+};-\vec{P}_+)\, S(q_{n+-})\,
  \Gamma_\nu^{\rho(T/L)}(q_{n+};\vec{Q})\,\nonumber \\ 
&& \times  S(q_{n++}) \; \Gamma_\pi(k_{n-};\vec{P}_-)\; S(q_{n-})\,.
\end{eqnarray}
Use of \Eq{BSrhoTL} immediately shows that the longitudinal $\rho$ mode
cannot couple to $\pi\pi$.  The temperature dependence obtained for the
transverse $\rho$ coupling constant $g_{\rho \pi\pi}(T)$ is displayed in
Fig.~\ref{fig:grpp}.  Again one observes continuity with the
\mbox{$T=0$} result and a very weak $T$-dependence until about $0.8
T_c$.  Around $T_c$, the coupling constant decreases significantly.

For both interactions of the $\rho$ mode, and below $T_c$, one expects
qualitatively similar behavior from spatial modes associated with higher
meson Matsubara frequencies \mbox{$\Omega_m \neq 0$} except that the
effect of $O(4)$ symmetry breaking will be evident earlier.  Above
$T_c$, each $2\pi T$ increment to the meson Matsubara frequency adds
significantly to the quark effective mass in integrals like
\Eq{rhophoton}.  Since dynamical chiral symmetry breaking is now absent, 
the quark propagators are of Dirac vector character, the meson Matsubara
frequencies are the largest mass scale in the system, and  perturbative 
behavior becomes increasingly dominant.  
The various spatial modes from \mbox{$\Omega_m \neq 0$} are
characterized by masses much greater than that of the lowest mode
considered here.  We anticipate that this lowest mode
characterizes the qualitative behavior of the physical decay processes 
\mbox{$\rho^0 \to e^+\,e^-$} and \mbox{$\rho \to \pi\pi$}.
Certainly, a high temperature limit in which some modes vanish and 
others diverge would indicate a non-analytic behavior in the 
reconstructed physical amplitude that is physically untenable.   
We therefore estimate the relevant decay widths by combining the present 
results for the coupling constants with the relevant phase space factors
generated also from the lowest spatial mass modes.  
\begin{figure}[ht]
\centerline{
\epsfig{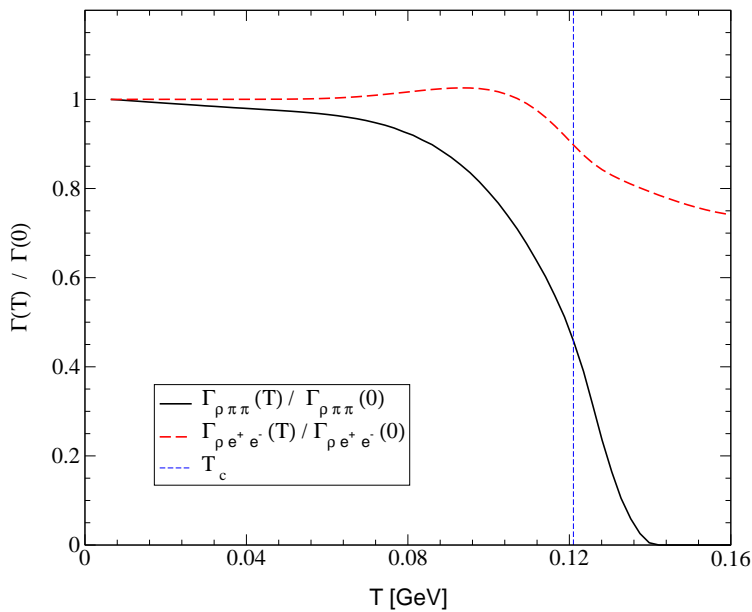}
}
\caption{\label{fig:rhowid} $T$-dependence of the transverse $\rho$ partial
widths due to electromagnetic \mbox{$e^+e^-$} decay (dashed line) and
strong \mbox{$\pi \pi$} decay (solid line)  corresponding to the
coupling constants shown in Fig.~\protect\ref{fig:grpp}.  }
\end{figure}

This leads to the electromagnetic decay width  
\begin{eqnarray}
        \Gamma_{\rho^0 \rightarrow e^+\,e^-}(T) &=& 
                \frac{4\pi\,\alpha^2\,M^T_\rho(T)}{3\;g_\rho(T)^2}~,
\end{eqnarray}
while the corresponding  strong decay width is
\beq
\Gamma_{\rho\rightarrow\pi\pi}(T) = \frac{g_{\rho\pi\pi}(T)^2}{4\pi}
\frac{M_\rho^T(T)}{12}\left[1-\frac{4M_\pi(T)^2}{M_\rho^T(T)^2}
                                                      \right]^{3/2}.
\label{rhowidth}
\eeq
The $T$-dependence estimated in this way for the decay widths is due to
the response of the quark substructure to the heat bath, particularly
the restoration of chiral symmetry.  The results are displayed in
Fig.~\ref{fig:rhowid}.  The contrast between the behavior of the
electromagnetic and strong widths near and just above $T_c$ should be a
more robust finding than the details of the individual processes.  The
strong width decreases rapidly and vanishes just above $T_c$ while the
electromagnetic width remains within 20\% of the \mbox{$T=0$} value.
Part of the strong decrease of the intrinsic $\pi\pi$ width of the
transverse $\rho$ is due to the decrease in the coupling constant,
however the dominant effect is the $T$-dependence of the last factor in
\Eq{rhowidth}.  As displayed in  Fig.~\ref{fig:rhomass},
$2 M_\pi(T)$ rises faster with $T$ than does $M_\rho^T(T)$ until at  
\mbox{$T=1.17~T_c$} we have \mbox{$M_\rho^T = 2 M_\pi$}.   Beyond this 
point, the phase space factor vanishes and the strong  decay 
\mbox{$\rho^T \to \pi\pi$} is blocked.   This suggests that the total 
$\rho^T$ width of 
151~MeV at $T=0$ decreases by about 50\% near \mbox{$T=T_c$} and drops 
sharply to the electromagnetic value of about 6~keV by \mbox{$T=1.17~T_c$}.
 
This narrowing of the intrinsic decay width of the vector meson mode in
the heat bath is a mechanism that is distinct from the collisional 
broadening effect~\cite{RW99} from the many-hadron environment.   The 
present work indicates that there is a non-trivial $T$-dependence to 
intrinsic coupling constants such as $g_{\rho \pi \pi}$ and decay phase
space.   The intrinsic effect tends to significantly decrease the decay 
width; the many-hadron medium effects have the opposite influence.   
Phenomenological forms for the $T$-dependence of $M_\rho(T)$ and 
\mbox{$\Gamma_{\rho^0 \to \pi \pi}(T)$} have often been explored  in 
studies of medium effects in heavy-ion collisions.    For  example, 
both an increase in the width  of the form
\mbox{$\Gamma_\rho^I /(1-T^2/T_c^2)$}, and a $\rho$ mass decreasing by 
50\% at $T_c$  have been explored in an effort to understand the
heavy-ion dilepton spectrum ~\cite{sb}.  A coordinated approach is called
for in which
hadronic collisional broadening mechanisms are built upon intrinsic
coupling constants that respect the temperature and density dependence of the 
quark-gluon content.

\section{Behavior at large $T$}

\begin{figure}[ht]
\centerline{
\epsfig{figure=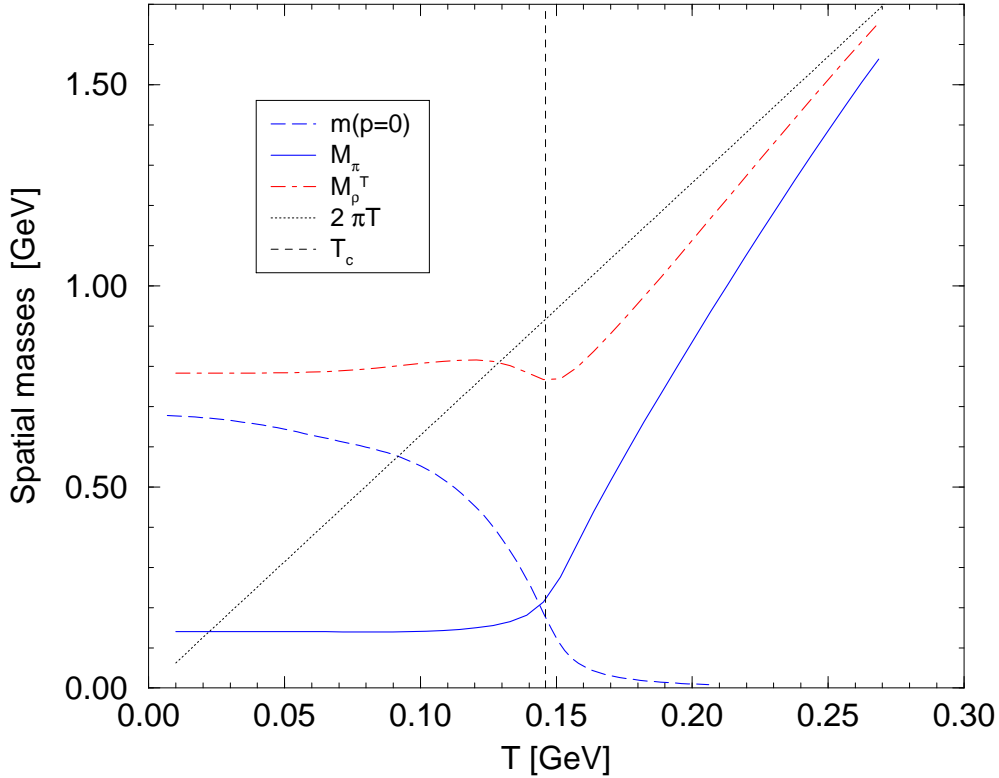,height=10.5cm,angle=-90}
}
\caption{\label{fig:hiTmass} High $T$ behavior of the spatial masses of 
the $\pi$ and transverse $\rho$ modes from the rank-1 separable model.}
\end{figure}
The quark deconfinement point $T_d$ and the chiral restoration point
$T_c$ are generally expected be to identical or nearly
so~\cite{laermann}.  (The present separable model produces
\mbox{$T_d=T_c$} in rank-1 and \mbox{$T_d=0.9~T_c$} in rank-2.)  It might
be expected therefore that above $T_c$ meson modes should dissolve in
favor of a gas of essentially massless quarks.  However for a
significant temperature range above $T_c$, the spatial $\pi$ and $\rho$
modes studied here continue to be stable against $\bar q q$ dissociation
and do not dissolve into a free quark gas.  This situation is clearest
for the simpler rank-1 separable interaction.  The results for
$M_\pi(T)$ and $M_\rho^T(T)$ obtained from the eigenmass condition
\mbox{$\lambda(-M^2) =1$} are displayed in \Fig{fig:hiTmass} along with
the quark dynamical mass function at \mbox{$p=0$}.  The masses of both
spatial meson modes approach the asymptotic behavior $2\pi T$ from
below.  This asymptotic behavior has been discussed
previously~\cite{gocksch} and is observed in lattice
simulations~\cite{laermann,gocksch}.

The manner in which the $2\pi T$ behavior at large $T$ emerges from the
present description is illustrated well by the rank-1 model.  For
\mbox{$T>T_c$}, the dynamically generated mass function of the quarks is
essentially negligible, and the quark propagator is dominated by the Dirac 
vector amplitude \mbox{$\sigma_V(q_n^2) \sim 1/q_n^2$}.  For the spatial 
$\pi$ mode characterized by \mbox{$P=(0,\vec{P})$}, the loop integral for 
the ``polarization'' function or BSE eigenvalue $\lambda_\pi(\vec{P}^2)$, 
given by \Eq{lampi}, yields at large $T$ 
\begin{equation}
\lambda_\pi(\vec{P}^2) \approx 
\frac{16 D_0}{3}\,  T  \int \frac{d^3q}{(2\pi)^3} \,
f_0^2(\pi^2T^2 +\vec{q}^{\,2}) 
\frac{ \pi^2T^2 + \vec{q}^{\,2}-\frac{\vec{P}^2}{4} }
{[\pi^2T^2 + (\vec{q}+\vec{P}/2)^2][\pi^2T^2 + (\vec{q}-\vec{P}/2)^2]} \;,
\label{lampiT}
\end{equation}  
where only the dominant zeroth fermion Matsubara mode has been retained.
For \mbox{$T \gg \Lambda_0/\pi$}, with $\Lambda_0$ being the range of
the interaction form factor $f_0$, only small $q$ is relevant and the
position of the lowest singularity as a function of \mbox{$\vec{P}^2
<0$} approaches $2\pi T$.  The higher the temperature, the more
$\lambda_\pi(\vec{P}^2)$ is suppressed except near the singularity.  The
value \mbox{$\lambda_\pi(-M^2)=1$} must be encountered before the
divergence and thus \mbox{$M_\pi(T) \sim 2\pi T - \Delta_\pi(T)$} where
$\Delta_\pi$ is a positive mass defect that will typically decrease with
$T$.  Thus the spatial meson mass or screening mass will approach the
thermal mass of a pair of massless fermions from below.   This limit
has also been demonstrated from the pseudoscalar correlator within the 
Nambu--Jona-Lasinio model~\cite{FF94}.

In the general case, the detailed temperature behavior of the mass defect 
$\Delta_\pi(T)$, or the nature of the approach to the $2\pi T$ limit,  
depends upon the asymptotic behavior of the quark amplitudes $A(q^2)-1$ and 
$B(q^2)$.   In QCD, their leading asymptotic behavior is $\sim 1/q^2$ apart 
from slow logarithmic corrections.   The form factors chosen in this initial
work within a separable model induce an exponential fall-off.  It is to be 
expected therefore that the present model estimate of $\Delta_\pi(T)$
will decrease too rapidly with $T$.    

\begin{figure}[ht]
\centerline{
\epsfig{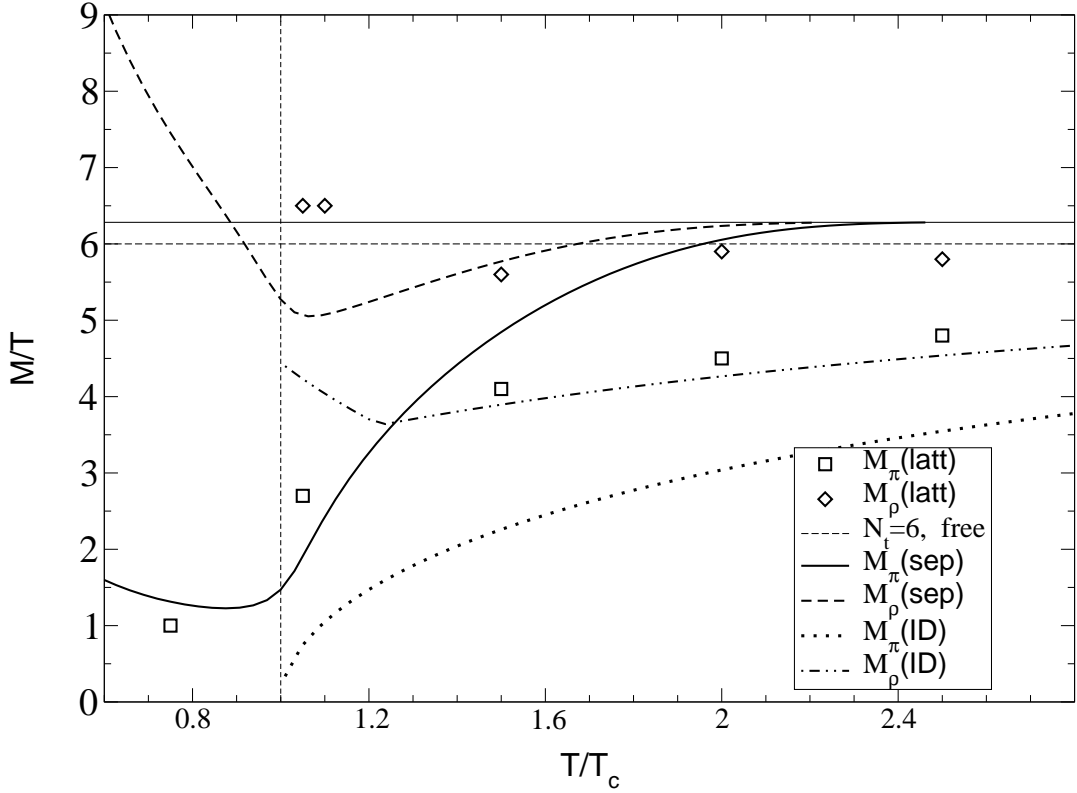}
}
\caption{\label{fig:latt_cf} Spatial masses from the rank-1 
separable model (sep) and the infrared-dominant model (ID) compared to 
spatial screening masses from lattice QCD simulations (latt) taken 
from \protect \cite{gocksch}.}
\end{figure}
In \Fig{fig:latt_cf} the $\pi$ and transverse $\rho$ spatial masses at
\mbox{$T>T_c$} from both the rank-1 separable model and the ID model 
(see the Appendix for more details on the ID model) are presented in
comparison with lattice QCD simulations of spatial screening
masses~\cite{gocksch}.  The solid horizontal line marks $2\pi$ while the
lower horizontal dot-dashed line represents the lattice free limit
corrected~\cite{gocksch} for the lattice time extent $N_t$.  From
\Fig{fig:latt_cf} it is evident that for the $\pi$ at \mbox{$T>T_c$},
the spatial or screening mass defect \mbox{$\Delta =$} \mbox{$2\pi
T-M(T)$} is decreasing more rapidly above $T_c$ in the rank-1 separable
model than is evident from the lattice simulations.  This is consistent
with the exponential behavior of the employed form factors.   It is
also consistent with the absence of  quark vector self-energy amplitudes 
$A(p,T)-1$ and $C(p,T)-1$ through which interactions can be quite 
persistent in the asymptotic region.    This can be demonstrated within
the chiral limit ID model where those amplitudes are strong and indeed
have the power law fall-off. As seen in \Fig{fig:latt_cf}, the resulting
mass defect for both $\pi$ and $\rho$ is in
fact too strong when compared with the lattice QCD simulations.  This
persistent self-interaction well above $T_c$, which slows the approach
to free behavior such as Stefan-Boltzmann thermodynamics~\cite{brs}, may
well be what is signaled by the lattice QCD data in \Fig{fig:latt_cf}.

\section{Discussion}

We have explored $\pi$ and $\rho$ spatial correlation modes at
\mbox{$T>0$} within the rainbow-ladder truncation of the quark
Dyson-Schwinger equation and the $\bar q q$ Bethe-Salpeter equation in
the Matsubara formalism.  With parameters fitted to \mbox{$T=0$}
properties, the model possesses dynamical chiral symmetry breaking and
quark confinement.  A simple separable form of the effective interaction
is employed and this facilitates the use of sufficient Matsubara modes to
allow coverage of low temperatures as well as the transition region.  
Deconfinement and
chiral restoration transition temperatures, $T_d$ and $T_c$, are very
similar and in the range 100-150~MeV.  The $T$-evolution of the $\pi$
and $\rho$ $\bar q q$ states in the presence of the deconfinement and
chiral restoration mechanisms is studied.  The degree to which the model
respects the axial vector Ward-Takahashi relation is evaluated in terms
of the exact pion mass relation and the related GMOR relation.  The
$\rho$ electromagnetic coupling constant $g_\rho$ and the strong
coupling constant $g_{\rho \pi\pi}$ are also obtained as a function of
$T$.  Estimates are made for the $T$-dependence of the widths for
\mbox{$\rho^0 \to e^+e^-$} and \mbox{$\rho \to \pi \pi$}.   Finally,
the high $T$ behavior of the spatial masses is compared to that of
spatial screening masses from lattice QCD simulations~\cite{gocksch}.

The masses $M_\pi(T)$, $M_\rho^T(T)$ and $M_\rho^L(T)$ are found to be
almost $T$-independent below $T_c$ followed by a strong increase.  This
behavior is characteristic of lattice QCD
simulations~\cite{laermann,gocksch} and DSE studies~\cite{mrs,MRST00}.
Our estimate of the \mbox{$\rho \to \pi \pi$} strong decay width shows a
decrease with $T$ such that, by $T_c$, it has been reduced by 50\%.  Our
arguments suggest a phase-space blocking effect at about 25~MeV above
$T_c$.  (A similar phase-space blocking has been argued before only for
the strong $\pi\pi$ decay of the scalar-isoscalar partner of the pion
near $T_c$ ~\cite{scalar,MRST00}.)  The tendency here is for the
transverse $\rho$ mode above $T_c$ to be left with a narrow total width
typical of electromagnetic decay.  One would expect the mass of the
pseudoscalar $K$ correlation to rise with $T$ in a similar fashion to
$M_\pi$, while the masses of the vector $\phi$ and $K^\star$ modes
should rise like the $\rho$.  This suggests that the vector modes
$\rho$, $K^\star$ and $\phi$ tend to be trapped with their relatively
long electroweak lifetimes and with significantly increased masses for a
domain of high temperatures above the transition.  Between
0.5-1.0$~T_c$, the transition probability connecting vector correlations
with pairs of pseudoscalars would be reduced.  This suggests that within
the gas of pions and other pseudoscalars that dominate the hot hadronic
product from heavy-ion collisions, the role of vector meson correlations
in producing the dilepton spectra could be significantly less than
conventional expectations.  The present findings follow from the
response of the quark-gluon content of the mesons to the heat bath.  A
different phenomenon is the coupling of the meson modes to the
many-hadron environment which introduces a collisional broadening
effect~\cite{RW99}.  An approach that incorporates both phenomena is
clearly called for.

Only the spatial or screening $\bar q q$ masses have been investigated
within this model.  The temporal masses (sometimes called dynamical or
pole masses) provide a different characterization of the correlations.
Lattice simulations indicate that spatial masses become much larger than
the temporal masses above $T_c$~\cite{TARO99}.  A Nambu--Jona-Lasinio
model study~\cite{FF94} found that they are significantly different only
in the range 150~MeV~\mbox{$<T<$}~350~MeV.  A possible explanation for
this discrepancy lies in our finding that the mass defect
\mbox{$\Delta =$} \mbox{$2\pi T-M(T)$} at high $T$ is significantly 
influenced by residual non-perturbative interaction effects  in the Dirac
vector amplitude \mbox{$A(P^2)-1$} of the quark self-energy.   Such a 
term is not present in the Nambu--Jona-Lasinio model which produces a 
momentum-independent quark constituent mass.   The present model allows
low temperature confinement and a momentum-dependent quark self-energy
in a simple way but at a cost of an exponential asymptotic fall-off
instead of a more realistic power law fall-off.  Comparison with 
lattice QCD results illustrates the connection between asymptotic 
behavior of the interaction and the high $T$ behavior of the mass defects 
\mbox{$\Delta_\pi(T)$} and \mbox{$\Delta_\rho(T)$}.

The simplicity of a separable  representation of the effective
quark-quark  interaction of the type studied here might be of advantage 
in the consideration of pion loop effects and bulk thermodynamic properties 
of the hadron-quark matter phase transitions.   Initial thermodynamic
considerations have been reported recently~\cite{bt}.  The only
$T$-dependence of the effective quark-quark interaction implemented by
the present separable model is that generated by the Matsubara
frequencies that enter through the momentum dependence.  The strength
and range parameters have been kept $T$-independent and no attempt has
been made to introduce explicit $T$-dependent characteristics such as a
Debye mass.  Such considerations are more appropriately handled in
approaches that have a better connection to a perturbative gluon
propagator~\cite{MRST00}.

\nonumsection{Acknowledgements}
\noindent
We acknowledge fruitful interactions and conversations with B. Van den
Bossche, M. Buballa, C.D. Roberts, and S. Schmidt. The work of G.B.  and
Y.L.K. has been supported by the Max-Planck-Gesellschaft and by the DFG
Graduiertenkolleg ``Stark korrelierte Vielteilchensysteme''.  D.B. and
G.B. gratefully acknowledge financial support by the Deutscher
Akademischer Austauschdienst (DAAD) for visits to the Center for Nuclear
Research at Kent State University where part of this work was conducted.
Y.L.K. also acknowledges the Russian Fund for Fundamental Research,
under contract number 97-01-01040, and the support of the
Heisenberg-Landau program.  P.C.T. and P.M. acknowledge support by the
National Science Foundation under Grant Nos. INT-9603385 and PHY97-22429
and the hospitality of the University of Rostock where part of this work
was conducted during several visits.

\appendix

\noindent
The extension to \mbox{$T>0$} of the ID model introduced at $T=0$ by
Munczek and Nemirovsky~\cite{mn} provides a semi-analytic perspective
on the large $T$
behavior of the meson masses.  For the Feynman-like gauge
used in the present study, the effective interaction of this model is
specified by
\beq
D(p-q) \to  (2\pi)^4 \frac{3 \eta^2}{16} \delta^4(p-q) \; , 
\label{DMN}
\eeq
which is to be used in the DSEs given in \Eqs{dseA} and (\ref{dseB}) and
also in the BSE given in \Eq{bs}\footnote{The results are unchanged when
Landau gauge is used if \Eq{DMN} is scaled up by $4/3$.}.  In the chiral
limit, closed form expressions exist for the resulting quark propagator
amplitudes and for the ladder BSE eigenvalues $\lambda(P^2)$ in the
pseudoscalar and vector channels of interest here.  The chiral limit DSE
solution is of the general form given in \Eq{prop} with
\begin{eqnarray}
\sigma_A(p^{2})=\sigma_C(p^{2}) & = & \left\{
\begin{array}{cl}
\frac{2}{\eta^2}, & p^{2}\leq \frac{\eta ^{2}}{4} \\
\frac{2}{p^2}\left[ 1+\left(1+\frac{2\eta ^{2}}{p^{2}}\right)
^{\frac{1}{2}}\right]^{-1} , & p^{2}\geq \frac{\eta ^{2}}{4}
\end{array}\right. \nonumber \\ & &  \nonumber \\
\sigma_B(p^{2}) & = & \left\{
\begin{array}{cl}
\; \; \; \frac{1}{\eta^2}(\eta ^{2}-4p^{2})^{\frac{1}{2}}, &
\; \; \; \; \; \; \; \; \; p^{2}\leq \frac{\eta ^{2}}{4} \\
0, & \; \; \; \; \; \; \; \; \; p^{2}\geq \frac{\eta ^{2}}{4}
\end{array}\right. . \label{MNsoln}
\end{eqnarray}
The quark mass function is \mbox{$m^2(p^2)=\eta^2/4-p^2$} for 
\mbox{$p^2\leq \eta ^2/4$}, and \mbox{$m^2(p^2)= 0$} otherwise.  There
is no mass-shell, i.e. \mbox{$p^2 + m^2(p^2) \neq 0$} for any $p^2$, and
there is quark confinement.  The above solution holds at \mbox{$T=0$}
and at \mbox{$T>0$} with $p^2$ replaced by 
\mbox{$p_n^2=$} \mbox{$\omega_n^2 + \vec{p}^{\,2}$}.  Thus for 
\mbox{$T>T_c=$} \mbox{$\eta/(2\pi)$} there is chiral restoration 
(\mbox{$\sigma_B=0$}).  
Results from this model at finite $T,\mu$ have been obtained for quark
thermodynamics outside the phase boundary of chiral restoration as
governed by the quark propagator~\cite{brs}.  The behavior of the masses
of both the $\pi$ and $\rho$ modes for \mbox{$T< T_c$} (and also for
chemical potential dependence for \mbox{$\mu < \mu_c$}) have been
obtained previously~\cite{mrs}.

With \Eq{DMN}, the BSE given in \Eq{bs} becomes
\begin{equation}
\lambda(P^2) \Gamma(q;P) 
= -\frac{\eta^2}{4} \, \gamma_\mu S(q_+) \Gamma(q;P)S(q_-) \gamma_\mu \;,
\label{MNbs}
\end{equation}
and solutions are possible if $q$ is fixed by $P$.  At \mbox{$T>0$}, the
solution that connects smoothly to the \mbox{$T=0$} solution has the
normally independent variable \mbox{$q_n = (\omega_n, \vec{q})$}
restricted to \mbox{$q_n \to (\omega_0, \vec{0})$}.  To obtain spatial
meson modes to compare with the separable model results, we again set
\mbox{$P = (0, \vec{P})$}.   Within the temperature domain 
\mbox{$T<T_c=\eta/(2\pi)$}, the results for the pseudoscalar and vector 
spatial modes are particularly simple and have been discussed
previously~\cite{mrs}.  One finds \mbox{$\lambda_\pi(0)=1$} for the
chiral limit $\pi$; thus \mbox{$M_\pi=0$}.  For the vector meson one
finds that \mbox{$\lambda_\rho^T(-\eta^2/2)=1$}; thus
\mbox{$M_\rho^T=\eta/\sqrt{2}$}.  These are also the correct \mbox{$T=0$} 
results of the model.  They hold over the finite temperature domain for
which the quark mass function appropriate to the propagators occurring
in the BS \Eq{MNbs} is nonzero.  For the equation appropriate to a meson
of mass $M$, the relevant domain for the present model is \mbox{$\eta^2
-4s >0$} where \mbox{$s=\pi^2\, T^2 -M^2/4$}.  For \mbox{$M_\pi=0$},
this temperature domain corresponds to that for which the model DSE
generates a dynamical quark mass function in accord with the Goldstone
theorem.  For larger $M$, this temperature domain will be larger.  The
vector result \mbox{$M_\rho=\eta/\sqrt{2}$} holds for the larger
temperature domain \mbox{$T<\sqrt{3/2} T_c$}.  We fix the single
parameter \mbox{$\eta=1.107$}~GeV, so that \mbox{$M_\rho=0.783$}~GeV.
This produces \mbox{$T_c=0.176$}~GeV.

Beyond the temperature domain where the meson mass is constant, one
finds \mbox{$\lambda_\pi(\vec{P}^2) = $} 
\mbox{$\eta^2 \, s_- \, \sigma_A(s_+)^2$}, where \mbox{$s_\pm = $}
\mbox{$\pi^2 T^2 \pm \vec{P}/4$}, and the $\rho$ eigenvalue is
simply given by \mbox{$\lambda_\rho^T(\vec{P}^2) = $}
\mbox{$\frac{1}{2}\, \lambda_\pi(\vec{P}^2)$}.
In general both functions $\lambda(-M^2)$ decrease with increasing $T$
and increase with increasing $M$.  Thus the spatial eigenmode condition
\mbox{$\lambda(-M^2)=1$} will tend to maintain \mbox{$M_\rho^T(T) > M_\pi(T)$}
while both masses rise with $T$, and the obtained behaviour of the
masses $M_\rho^T$ and $M_\pi$ is qualitatively the same as what we found
in the separable model.   The main difference is that the mass defect
\mbox{$\Delta(T)=2 \pi T - M(T)$} is significantly larger in the ID
model.  The leading large $T$ behavior
\beq
M_\rho^T(T)^2 \to (2\pi T)^2 \, \left( 1 - \frac{\eta}{\pi T} 
                          \cdots \right) \; ,
\label{asymM} 
\eeq
leads to the spatial mass defect of the transverse $\rho$ mode
having the asymptotic value \mbox{$\Delta_\rho(T) \to \eta$}.  
\Fig{fig:latt_cf} displays the behavior for both $\pi$ and $\rho$.

\nonumsection{References}

%
%
%
%
%
\end{document}